\documentclass[aps,twocolumn,superscriptaddress,letterpaper]{revtex4}
\usepackage{graphicx,amssymb,bm,natbib,amsfonts,amsmath}
\usepackage{hyperref}

\newcommand{\kpar}{{\bf k}$_\parallel$}
\newcommand{\kz}{k$_z$}

\begin{document}

\title{Coexistence of Sharp Quasiparticle Dispersions and Disorder Features in Graphite}

\author{S.Y. Zhou}
\affiliation{Department of Physics, University of California,
Berkeley, CA 94720, USA}
\author{G.-H. Gweon}
\affiliation{Department of Physics, University of California,
Berkeley, CA 94720, USA} 
\affiliation{Materials Sciences Division,
Lawrence Berkeley National Laboratory, Berkeley, CA 94720, USA}
\author{C.D. Spataru}
\affiliation{Department of Physics, University of California,
Berkeley, CA 94720, USA}
\affiliation{Chemical Sciences Division,
Lawrence Berkeley National Laboratory, Berkeley, CA 94720, USA}
\author{J. Graf}
\affiliation{Materials Sciences Division, Lawrence Berkeley National
Laboratory, Berkeley, CA 94720, USA}
\author{D.-H. Lee}
\affiliation{Department of Physics, University of California,
Berkeley, CA 94720, USA}
\affiliation{Materials Sciences Division,
Lawrence Berkeley National Laboratory, Berkeley, CA 94720, USA}
\author{Steven G. Louie}
\affiliation{Department of Physics, University of California,
Berkeley, CA 94720, USA}
\affiliation{Materials Sciences Division,
Lawrence Berkeley National Laboratory, Berkeley, CA 94720, USA}
\author{A. Lanzara}
\affiliation{Department of Physics, University of California,
Berkeley, CA 94720, USA}
\affiliation{Materials Sciences Division,
Lawrence Berkeley National Laboratory, Berkeley, CA 94720, USA}

\date{February 28, 2005}

\begin{abstract}
Angle resolved photoelectron spectroscopy (ARPES) on azimuthally disordered graphite demonstrates that sharp quasiparticle dispersions along the radial direction can coexist with a complete lack of dispersion along the azimuthal direction. This paradoxical coexistence can be explained in terms of van Hove singularities in the angular density of states. In addition, non-dispersive features at the energies of band maxima and saddle points are observed and possible explanations are discussed. This work opens a new possibility of studying the electronic structure of novel layered materials using ARPES even when large single crystals are difficult to obtain. 
\end{abstract}


\maketitle

The ability to sharply resolve crystal momentum values of single particle excitations has made ARPES a very powerful tool in addressing the electronic structure of solid, as has been successfully demonstrated on single crystalline samples over the past decades \cite{Hufner}.  Due to the translational symmetry along the surface of a single crystal, the crystal momentum parallel to the surface (\kpar) is conserved during the photoemission process, allowing a complete momentum space map of the initial state.  This holds despite the short photoelectron lifetime  \cite{Hedin,Smith,Lindroos} which can severely broaden the resolution of the momentum perpendicular to the surface (\kz).  Indeed, even in the limit of an extreme \kz~broadening that results in no resolution of \kz, strong ARPES dispersions are expected as a function of \kpar, since the one dimensional density of states (1D-DOS) D$_z$(E) $\propto$ d\kz/dE obtained by integrating over \kz~is dominated by contributions from van Hove singularities in high symmetry planes \cite{Lindroos}.

On the contrary, for those systems characterized by orientationally disordered domains, i.e.~polycrystalline materials, the translational symmetry is preserved only within each domain.  As a consequence, the dispersion measured by ARPES is the average dispersion over different domains, or equivalently azimuthal angle $\phi$, which in general leads to no dispersion. However, extending the 1D-DOS D$_z$(E) scenario for \kz~discussed above further to the plane, there is an interesting possibility that a layered polycrystalline sample, with a strong  azimuthal disorder, can nevertheless give distinct dispersions in the radial direction.  This would happen if the average dispersion is dominated by those along the high symmetry directions due to van Hove singularities in the angular density of states D$_\phi$(E) $\propto$ d$\phi/$dE.  To date, this possibility has never been demonstrated experimentally and photoemission studies on disordered samples have focused on angle-integrated features without any momentum information.

In this paper, we report a high resolution ARPES study on the electronic structure of azimuthally disordered graphite.  For the first time, we report clear evidence that sharp quasiparticle dispersions, in agreement with band structure calculation along the high symmetry directions, can coexist with a circular Fermi energy intensity map, a definitive signature of azimuthal disorder \cite{Santoni}.  In addition we report non-dispersive features at the energies of band maxima or saddle points, which are attributed to the loss of momentum information by indirect photoemission process or elastic scattering. A practical implication of this study is that more ARPES opportunities can be made available for layered materials even when high quality single crystals of large size are difficult to obtain. 

ARPES data were collected at beam line 10.0.1 of the Advanced Light Source (ALS) at the Lawrence Berkeley National Laboratory, using an SES-R4000 analyzer.  The wide angular mode with acceptance angle of 30$^{\circ}$ and angular resolution of 0.9$^{\circ}$ was utilized for most scans, while high resolution angular mode with acceptance angle of 14$^{\circ}$ and angular resolution of 0.1$^{\circ}$ was utilized for one scan.  The total instrumental energy resolution was 15 meV at 25 eV photon energy and 25 meV for other photon energies used (40, 55, 60 eV).  The sample used was a grade ZYA highly oriented pyrolytic graphite (HOPG), obtained commercially from Structure Probe Inc.  The sample was cleaved {\em in situ} in an ultra high vacuum better than 1.0$\times$10$^{-10}$ Torr and measured at temperature 50 K.

\begin{figure*}[t]
\includegraphics[width=16cm]{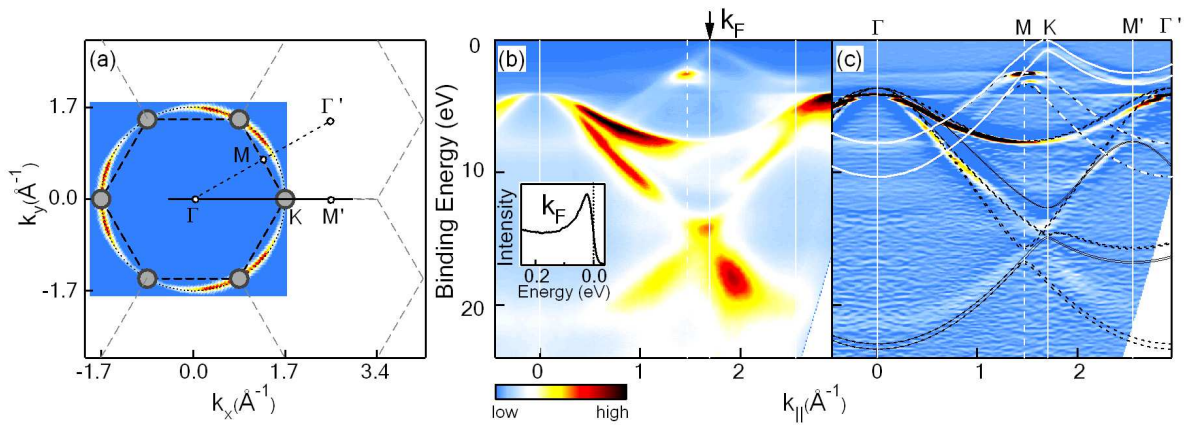}
\label{fig1}
\caption{(a) Fermi energy intensity map. The hexagonal Brillouin zone (dashed lines) and Fermi surface (shaded circles) expected for single crystalline graphite are drawn schematically. (b) Intensity map versus binding energy and in-plane momentum along the solid line in (a) taken at 60 eV photon energy. Arrow marks the Fermi energy crossing point (k$_F$). The inset shows EDC at k$_F$ taken at 25 eV photon energy in the high angular resolution (0.1$^{\circ}$) mode. (c) Second derivative of raw data in (b) with respect to energy. LDA dispersions along both $\Gamma$-K-M' direction (solid lines) and $\Gamma$-M-$\Gamma$' direction (dashed lines) are plotted for comparison. The Brillouin zones are labeled on top of this panel for the two high symmetry directions.}
\end{figure*}

Figure 1(a) shows an ARPES intensity map measured at the Fermi energy taken at 40 eV photon energy.  Throughout this paper, we use a color scale such that black represents high intensity in the raw data and blue represents low intensity.  According to band structure calculation, a constant \kz~cross section of graphite Fermi surface can be a small hole pocket, a small electron pocket, or a point located at the six corners of the hexagonal Brillouin zone (dashed line in figure 1(a)), depending on the value of \kz~\cite{BandStructure, el-h}.  Experimentally, the predicted small electron or hole pockets have yet to be resolved, and measurements on single crystalline samples have shown only small dots of high intensity at these corners \cite{Santoni}, schematically drawn as shaded circles in Figure 1(a). For the  graphite sample under study, the Fermi energy intensity map, symmetrized by three fold rotations to fill the entire Brillouin zone, shows a perfectly circular pattern  \footnote {The intensity variation along the circle is attributed to photoemission matrix element and is not a major concern here.}, in contrast to what is expected for single crystalline graphite. This is attributed to the angular spread of the dots to a circle due to the azimuthal disorder of the sample \cite{Santoni}. 

Figure 1(b) shows an ARPES intensity map as a function of binding energy and in-plane momentum \kpar, corresponding to the momentum cut shown as a solid line in Figure 1(a).  Despite the strong azimuthal disorder giving a circular Fermi energy intensity map in Figure 1(a), we observe, surprisingly, very clear dispersions over the entire energy range.  Furthermore, at the Fermi energy crossing point k$_{F}$, a sharp coherent quasi-particle peak is observed.  This is shown in the inset, where an energy distribution curve (EDC), energy cut at a constant momentum, is plotted.  Here the half width of the EDC peak is 20 meV (50 meV FWHM due to the asymmetry of the line shape), defining the sharpest peak observed in graphite so far \cite{Kihlgren}.

In Figure 1(c) we report the second derivative of the raw data of Figure 1(b) with respect to energy.  The second derivative method has been used in the literature to enhance the direct view of the ARPES dispersion.  Local density approximation (LDA) band dispersions along two high symmetry directions $\Gamma$-K-M$^{'}$ (solid lines) and $\Gamma$-M-$\Gamma^{'}$ (dashed lines) are plotted in the same figure for a direct comparison. Despite a polycrystal-like sample implied by the Fermi energy intensity map, an excellent agreement is observed between the experiment and the theory.  We can identify the dispersions between 4 eV and 23 eV  as originating from the $sp^{2}$ orbitals with strong intra-layer $\sigma$ bonding (black lines), and the dispersions between Fermi energy and 11 eV as originating from the $p_{z}$ orbitals with weaker $\pi$ bonding (white lines).  We note that the calculated dispersions were stretched by 20$\%$ in energy throughout this paper, as suggested in the literature \cite{Bandwidening, Kihlgren, Louie}. The stretching of the LDA band dispersions is attributed to missing self-energy corrections in LDA, since {\em ab initio} quasiparticle calculations based on the GW method show that for graphite the quasiparticle band dispersion near the Fermi level is 15\% larger\cite{Louie}.

\begin{figure*}
\includegraphics[width=0.8\textwidth]{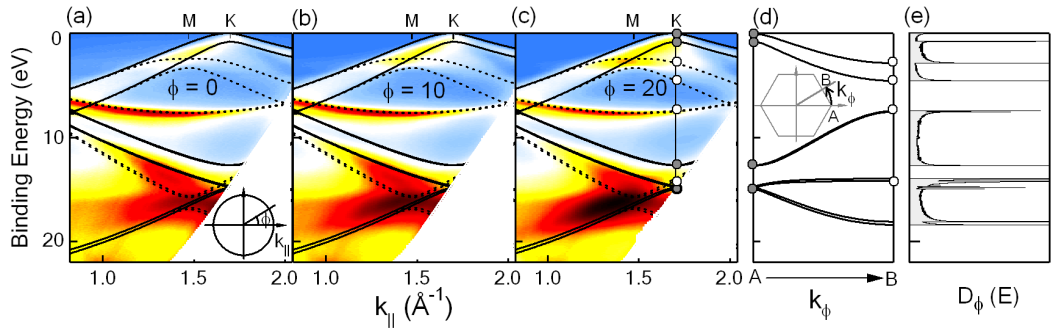}
\label{fig2}
\caption{Dispersions for azimuthal angles $\phi$=0$^{\circ}$(panel a); 10$^{\circ}$(panel b) and  20$^{\circ}$(panel c). The $\phi$ angle is defined in the inset of panel a.  LDA band dispersions along $\Gamma$-K-M' (solid lines) and $\Gamma$-M-$\Gamma$' (dotted lines) are plotted for comparison. (d) Calculated dispersions for single crystalline graphite along an arc (shown in the inset) with radius equal to $\Gamma$K distance. (e) Calculated density of states D$_\phi$(E) for single crystalline graphite by integrating over an arc from A to B (see inset of panel d). Singular peaks in D$_\phi$(E) occur at energies corresponding to band energy extrema, some of which are shown in panel(c,d) as shaded circles and open circles for A ($\Gamma$K direction) and B ($\Gamma$M direction) respectively.}
\end{figure*}

The direct comparison between Figure 1(a) and Figure 1(b,c) shows an apparent paradox in our data, namely, the coexistence of azimuthal disorder feature (Figure 1(a)) with single crystalline features (Figure 1(b,c)). This can be readily understood if we consider an angular average of the calculated dispersions.  Such an angular average would be necessary if the sample consisted of many small single crystallites with strong azimuthal disorder. 

Figures 2(a-c) show ARPES cuts for three azimuthal angles, $\phi$=0$^{\circ}$, 10$^{\circ}$, 20$^{\circ}$.  A direct comparison between panels a, b and c shows no appreciable angular dependence of the dispersions, establishing that the azimuthally invariant electronic structure of Figure 1(a) at the Fermi energy extends to the entire band width.  Thus, these data strongly support the azimuthal disorder model described in the previous paragraph, and indicate that the ARPES data measured are actually a 1D-DOS D$_\phi$(E) along the azimuthal direction $\phi$, in analogy with the well-known 1D-DOS D$_z$(E) along the \kz~direction \cite{Lindroos}. As in the latter case, then, one would expect that van Hove singularities arising from states along the high symmetry directions to contribute dominantly, and this gives an explanation why the measured dispersions accurately reflect the dispersions along the two high symmetry directions. The 1D-like van Hove singularities arise from states along high symmetry lines because these states have zero group velocity along the arc with constant \kpar~magnitude.

In Figure 2(d, e) we show LDA calculations supporting this reasoning.  In panel d, we show the dispersion for single crystalline graphite along an arc from A to B with radius equal to $\Gamma$K distance.  Within this arc, the $\Gamma$K direction corresponds to point A and the $\Gamma$M direction to point B respectively.  As expected, extrema in the band dispersion occur at the two high symmetry directions, A (shaded circles) and B (open circles).  The calculated 1D-DOS D$_\phi$(E) over this arc, and thus over the entire azimuthal angle range by symmetry, is shown in panel e. One can see diverging 1D van Hove singularities as sharp peaks occurring at energies where bands cross points A and B, which completely dominate over other contributions.  This nicely explains why well-defined sharp peaks with large dispersions can be observed in this azimuthally disordered sample, despite the fact that the observed data come from averaging over all azimuthal directions.

The data presented so far can be summarized as showing well-defined dispersions along the radial direction with a complete lack of dispersion along the azimuthal direction.  Therefore, our data suggest that the graphite sample under study consists of finite size single crystalline grains much smaller than the analysis area ($\approx$ 100 $\mu m$) with a complete azimuthal disorder.  However, each grain is large enough to allow for highly dispersive quasiparticles to exist.  In addition, we have measured the dispersion perpendicular to the surface, \kz~dispersion, using photon energy range from 34 to 155 eV at beam line 12.0.1 of the ALS, with a perfect agreement with previous results \cite{Law}.  This indicates that the crystalline order remains coherent along this direction, i.e.~perpendicular to graphene layers, over a length scale larger than the probing depth of ARPES (order of 10 \AA).  

We now discuss additional features at 2.9 (e$_{1}$), 4.3 (e$_{2}$) and 7.8 eV (e$_{3}$) characterized by no dispersion at all.  These features can already be observed in Figure 1(b,c) taken at 60 eV photon energy, appearing as sharp extended horizontal lines at the same energies.  Figure 3 shows a detailed view of these same features observed at another photon energy of 55 eV.  In panel a, we show the first derivative of the ARPES intensity map with respect to energy. The first derivative enhances rising or falling edges in the data, and thus is particularly useful for detecting non-dispersive peaks and edges. The energies for these non-dispersive features are marked by arrows on the right of this panel.  In panel b, we show EDCs at $\Gamma$ point (thin line) and \kpar$= 0.4 $ \AA$^{-1}$ (thick line). In these panels, in addition to highly dispersive features that we already discussed, one can clearly see the non-dispersive features, i.e. peaks at energies e$_{1}$ and e$_{3}$ and edge at energy e$_2$. The non-dispersive nature of these features can be checked more explicitly using EDCs.  For example, in panel c, a stack of EDCs over an
extended momentum range clearly show the non-dispersive nature of peak at e$_1$ and edge at e$_2$.  Such an analysis shows that the features at e$_1$, e$_2$ and e$_3$ are non-dispersive within $\approx$ 200 meV.

\begin{figure}
\includegraphics[width=7.5cm]{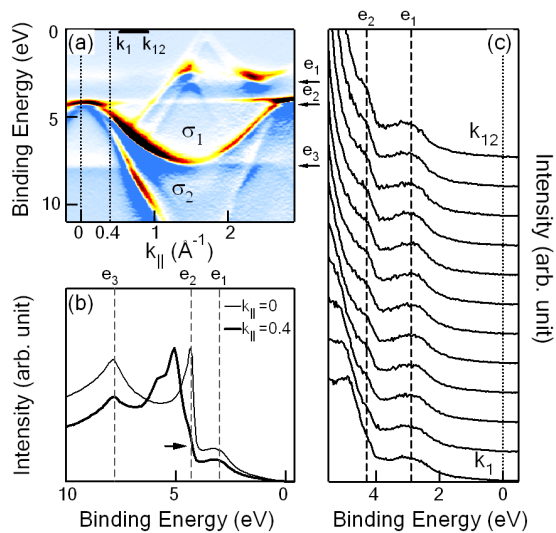}
\label{fig3}
\caption{(a) First derivative of raw data taken at 55 eV photon energy. (b) EDCs at $\Gamma$ point (thin line) and k$_{||}$= 0.4 $\AA^{-1}$ (thick line). The positions of these two cuts are shown as dotted lines in a. (c) EDCs for the momentum range indicated by a horizontal black line (from k$_1$ to k$_{12}$), marked in panel a.}
\end{figure}

Similar non-dispersive features have been noted and attributed to surface states or emission from isolated atoms  \cite{McGovern}. However, it is important to point out that in our data there is a strong connection between
these non-dispersive features and the highly dispersive features discussed above. Indeed, one can easily see from Figure 1(c) and Figure 3(a) that the energies of these non-dispersive features correspond to the band energy extrema.  Namely, e$_{1}$ corresponds to the saddle point of $\pi$ band, e$_{2}$ the degenerate maxima of $\sigma$$_{1}$ and $\sigma$$_{2}$ bands, and e$_{3}$ the saddle point of the $\sigma$$_{1}$ band. This strongly suggests that they are related to the bulk band structure rather than surface state. We suggest that these non-dispersive features can be explained by loss of electron momentum information in two possible ways, both of which are associated with the large density of states at the band energy extrema \cite{Bianconi}. One possibility is indirect transition, i.e.~non-k-conserving transition in photoemission process \cite{Law,Marchand}, for example, phonon assisted transition.  Another possibility is elastic scattering of electrons in either the initial state or the final state by inhomogeneity or disorder. In both these scenarios, the observed non-dispersive features correspond to 
density of states of graphite.  This explanation is consistent with the EDC line shape observed in panels (b,c).  That is, the non-dispersive feature at e$_2$ is sharp at low binding energy side because e$_2$ is the energy maximum of $\sigma$$_{1}$ and $\sigma$$_{2}$ bands, whereas the non-dispersive features at e$_1$ and e$_3$ are broad on both sides, because e$_1$ and e$_3$ are saddle point energies. We note that similar non-dispersive features are also observed in single crystalline graphite \cite{Law,Strocov}, which suggests that point defects definitely play an important role, in addition to the possible role by the azimuthal disorder we noted above. Interestingly, similar non-dispersive features are also observed in NaMo$_6$O$_{17}$ \cite{Gweon} and high T$_c$ superconductor \cite{Kaminski}. Future studies of these features may be interesting, in view of the general lack of understanding of the role of inhomogeneity or disorder in ARPES, and the importance of this topic in complex materials.

In conclusion, the data presented here provide a novel example of the coexistence of sharp quasiparticle dispersions with disorder features. Also, several non-dispersive features are identified at band energy extrema. Our study brings up an interesting possibility that layered crystals with azimuthal disorder can be studied with ARPES to obtain useful information. 

\begin{acknowledgments}
We thank C.M. Jozwiak, A. Bill, K. McElroy, and D.R. Garcia for useful discussions and A.V. Fedorov for experimental assistance at beam line 12.0.1. This work was supported by the Director, Office of Science, Office of Basic Energy Sciences, Division of Materials Sciences and Engineering of the U.S Department of Energy under Contract No.~DEAC03-76SF00098;  by the National Science Foundation through Grant No.~DMR03-49361 and Grant No.~DMR04-39768.  Support by the Sloan Foundation, the Hellman Foundation and computer resource at the NSF NPACI Center is also acknowledged.
\end{acknowledgments}

\end{document}